\documentclass[aps,prl,twocolumn,superscriptaddress,english]{revtex4}
\usepackage{amssymb}
\usepackage{graphicx}
\usepackage{amsmath}
\usepackage{soul}
\usepackage{amsthm}
\usepackage{amsfonts}
\usepackage[T1]{fontenc}
\usepackage[latin9]{inputenc}
\usepackage{array}
\usepackage{multirow}
\usepackage{color}
\usepackage{esint}
\usepackage{bm}
\usepackage{color}
\usepackage{bbm}
\usepackage{hyperref}
\usepackage{babel}
\usepackage{titlesec}

\begin{document}

\global\long\def\id{\mathbbm{1}}
\global\long\def\ui{\mathbbm{i}}
\global\long\def\ud{\mathrm{d}}

\title{Manipulating non-Hermitian skin effect via electric fields}
\author{Yi Peng}
\thanks{These authors contribute equally to this work.}
\affiliation{Shenzhen Institute for Quantum Science and Engineering,
Southern University of Science and Technology, Shenzhen 518055, China}
\affiliation{International Quantum Academy, Shenzhen 518048, China}
\affiliation{Guangdong Provincial Key Laboratory of Quantum Science and Engineering, Southern University of Science and Technology, Shenzhen 518055, China}
\author{Jianwen Jie}
\thanks{These authors contribute equally to this work.}
\affiliation{Shenzhen Institute for Quantum Science and Engineering,
Southern University of Science and Technology, Shenzhen 518055, China}
\affiliation{International Quantum Academy, Shenzhen 518048, China}
\affiliation{Guangdong Provincial Key Laboratory of Quantum Science and Engineering, Southern University of Science and Technology, Shenzhen 518055, China}
\author{Dapeng Yu}
\affiliation{Shenzhen Institute for Quantum Science and Engineering,
Southern University of Science and Technology, Shenzhen 518055, China}
\affiliation{International Quantum Academy, Shenzhen 518048, China}
\affiliation{Guangdong Provincial Key Laboratory of Quantum Science and Engineering, Southern University of Science and Technology, Shenzhen 518055, China}
\author{Yucheng Wang}
\thanks{Corresponding author: wangyc3@sustech.edu.cn}
\affiliation{Shenzhen Institute for Quantum Science and Engineering,
Southern University of Science and Technology, Shenzhen 518055, China}
\affiliation{International Quantum Academy, Shenzhen 518048, China}
\affiliation{Guangdong Provincial Key Laboratory of Quantum Science and Engineering, Southern University of Science and Technology, Shenzhen 518055, China}
\begin{abstract}
In non-Hermitian systems, the phenomenon that the bulk-band eigenstates are accumulated at the boundaries of the systems under open boundary conditions is called non-Hermitian skin effect (NHSE), which is one of the most iconic and important features of a non-Hermitian system.
In this work, we investigate the fate of NHSE in the presence of electric fields by analytically calculating the dynamical evolution of an initial bulk state and numerically computing the spectral winding number, the distributions of eigenstates, as well as the dynamical evolutions.
We show abundant manipulation effects of dc and ac fields on the NHSE, and that the physical mechanism behind these effects is the interplay between the Stark localization, dynamic localization and the NHSE. In addition, the finite size analysis of the non-Hermitian system with a pure dc field shows the phenomenon of size-dependent NHSE. We further propose a scheme to realize the discussed model based on an electronic circuit. The results will help to deepen the understanding of NHSE and its manipulation.
\end{abstract}
\maketitle

{\em Introduction.---} Hermiticity of Hamiltonian has been regarded as a fundamental requirement in standard quantum mechanics, and it ensures the conservation of probability and limits energy-values to be real in isolated systems. However, many systems, such as the nonequilibrium and open systems with gain and loss, can be effectively described by non-Hermitian Hamiltonians. Especially in recent years, non-Hermitian physics has attracted widespread attention in both theory~\cite{Ashida2020,Bender,Hatano,Lee2016,Shen2018,Gong2018,Yao2018,Kunst2018,Yang2020,Lee2019,Kawabata2019,Longhi2019,Jiang2019,Liu2020,Longwen2021,Bergholtz2021} and experiment~\cite{Schomerus,Luo2019,Xiao2020,Weidemann2020,Helbig2020,XXZhang,Zhang2021,Yi2021}. Various unique features of non-Hermitian systems without any Hermitian counterparts have been revealed, such as exceptional points and rings~\cite{Debowski2001,Wiersig2014,Hu2017,Chen2017,Hodaei2017,Zhang2019,Miri2019,Xu2017,Cerjan2019,Gong2019}, enriched topological classifications~\cite{Ueda2019,Magnea,Wojcik2020,Li2021,Hu2021,Liu2019}, and non-Hermitian skin effect (NHSE)~\cite{Yao2018,Kunst2018,Lee2019,Helbig2020,Alvarez2018,Slager2020,Zensen,Yifei2020,Okuma2020,LinhuLi}. NHSE, namely that a majority of eigenstates are localized near the boundary under open boundary conditions (OBC), is one of the most iconic properties of non-Hermitian systems. It drastically reshapes the bulk-boundary correspondence principle and motivates the establishment of generalized Brillouin zone~\cite{Yao2018,Kunst2018,Yang2020}. The interplay between the NHSE and other fundamental phenomena (e.g., localization induced by external magnetic fields, defects, disorder and quasiperiodic potentials~\cite{Longhi2019,Jiang2019,Liu2020,Longwen2021,LinhuLi,Lu2021,Shao2021,Hughes2021,MBLUeda}) has also attracted widespread attentions recently. On the other hand, electric field can induce the Stark localization or dynamical localization, and is also a frequently used fundamental method to
manipulate other physical effects, since it is easily realized and controlled. However, the effect of electric fields on NHSE was not considered before.

Now we focus on how to manipulate NHSE by using electric fields.
If the NHSE can be fully suppressed in the modulation process, the non-Hermitian effect may be eliminated, and the system may have the conservation of probability and all the eigenvalues may become real, even though the non-Hermitian term remains. Thus, manipulating NHSE is helpful for deepening our understandings of non-Hermitian quantum mechanics and the differences between Hermitian and non-Hermitian physics. Moreover, mastering how to control NHSE, we will be able to obtain or remove it on demand. Therefore, manipulating NHSE also has practical significance. In this work, we want to address whether electric fields can manipulate NHSE, and furthermore, if they can, whether richer and more interesting physics and applications will emerge in light of this.

%Aiming to answer above questions, we investigate the manipulation effects of electric fields on NHSE in one dimensional (1D) non-Hermitian systems. We firstly analytically study {\color{blue}a particle's dynamical behavior, which} can reveal the existence or non-existence of the NHSE. Then we consider the effects of three types of electric fields on the NHSE analytically and numerically.

{\em Model and results.---} We consider a one dimensional non-Hermitian system with non-reciprocal hopping under
the influence of electric fields, and the Hamiltonian is written as
\begin{equation}\label{hams}
\hat{H}= \sum_n(J_L|n\rangle\langle n+1| + J_R|n+1\rangle\langle n|)+E(t)a\sum_nn|n\rangle\langle n|,
\end{equation}
where $|n\rangle$ is the Wannier state localized on the lattice site $n$, $J_L(J_R)$ represents the leftward (rightward) hopping amplitude, $a$ is the lattice constant, being set as $1$ throughout this work, and $E(t)=e\xi(t)$, with $e$ and $\xi(t)$ being the particle's charge and external electric field, respectively.

\begin{figure}
\hspace*{-0.1cm}
\centering
 \includegraphics[width=0.48\textwidth]{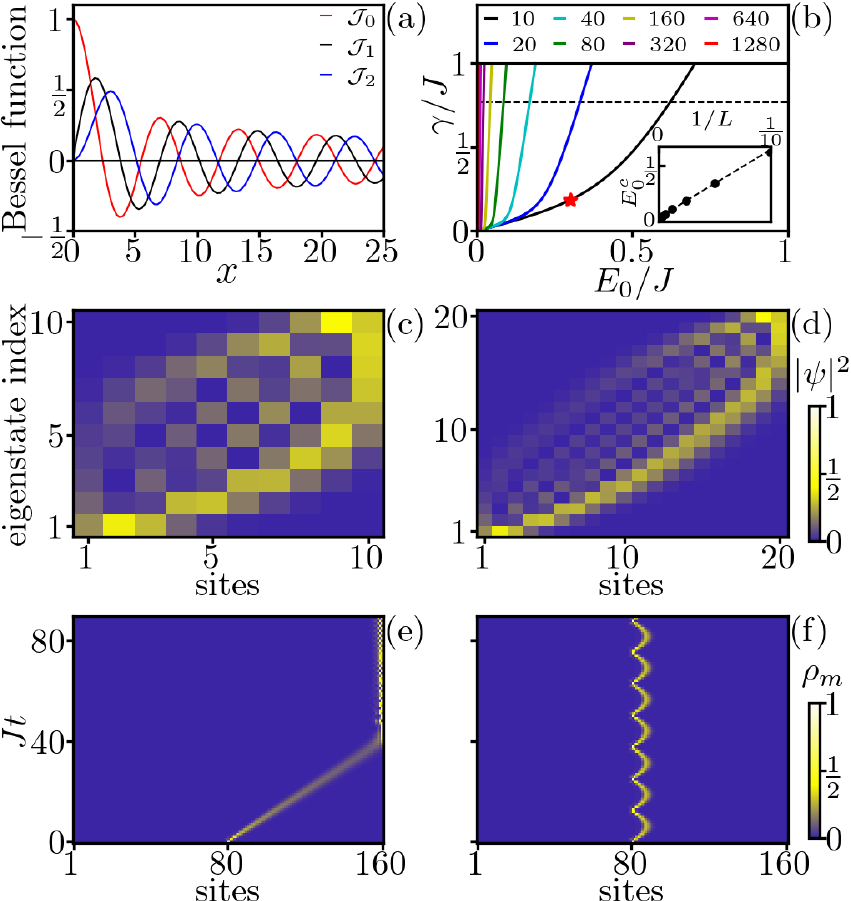}
\caption{\label{01}
(a) Distributions of the zeroth, first and second order Bessel function. We can see two characteristics used in the text: the amplitudes of oscillation decrease with increasing $x$ and when $x\rightarrow\infty$, $\mathcal{J}_m(x)\rightarrow 0$ for any $m$; $\mathcal{J}_0(0)=1$ and $\mathcal{J}_m(0)=0$ when $m\neq 0$, so we have $\mathcal{J}_m(0)=\delta_{m,0}$.
(b) Localization-delocalization transition characterized by winding number for finite lattices. The inset shows transition dc field strength $E_0^c$ versus $1/L$, given $\gamma=0.77$ marked as dashed line in its parent figure.
The distributions of eigenstates with (c) $L=10$ and
(d) $L=20$, and other parameters are $\gamma=0.185$ and $E_0=0.3$ as marked by red star in (b).
Dynamical evolution of a electron started from the lattice center
under (e) the weak dc field with $E_0=0.005$ and (f) the strong dc field with $E_0=0.5$, and other parameters are $L=160$, $\gamma=0.769$.}
\end{figure}

We can analytically confirm the existence or disappearance of NHSE by investigating the motion of a particle, this is because the particle initially localized in the bulk should move toward the boundary if the NHSE exists. We firstly substitute an arbitrary time-dependent quantum state $|\psi(t)\rangle=\sum_{m}C_{m}(t)|m\rangle$ into the Schr\"{o}dinger equation $i \partial_t |\psi(t)\rangle =\hat{H}(t)|\psi(t)\rangle$ and obtain the equation of motion for the time-dependent amplitudes $C_{m}(t)$,
\begin{eqnarray}\label{Seq}
i\partial_t C_{m}(t)=J_{L}C_{m+1}(t)+J_{R}C_{m-1}(t)+mE(t)C_{m}(t).
\end{eqnarray}
Here we set $\hbar=1$. By solving the Eq. (\ref{Seq}), for arbitrary $E(t)$, we can obtain the exact solutions,
\begin{widetext}
\begin{eqnarray}
C_{m}(t)&=&\sum_{n}(-1)^{m-n}C_{n}(0)e^{-i\eta(t)n}\mathcal{J}_{m-n}\left(2\sqrt{J_{L}J_{R}\left[\mathcal{U}^{2}(t)+\mathcal{V}^{2}(t)\right]}\right)\left[\frac{J_{R}}{J_{L}}\frac{i \mathcal{V}(t)-\mathcal{U}(t)}{i \mathcal{V}(t)+\mathcal{U}(t)}\right]^{\frac{m-n}{2}}.\label{main_fsol}
\end{eqnarray}
\end{widetext}
Here $\mathcal{J}_{m-n}(x)$ is the $(m-n)$th order Bessel function of the first kind, $\mathcal{U}(t)=\int_{0}^{t}\cos\left[\eta(t)-\eta(t')\right]dt'$ and
$\mathcal{V}(t)=\int_{0}^{t}\sin\left[\eta(t)-\eta(t')\right]dt'$ with
\begin{equation}\label{eta}
\eta(t)=\int_{0}^{t}E(t')dt'.
\end{equation}
This solution is valid for arbitrary initial bulk state $|\psi(0)\rangle$ and arbitrary $E(t)$, and the calculation details are in the Supplementary Materials~\cite{SM}. To simplify the expression, without loss of generality, we consider a specific initial state that only occupy a single Wannier lattice site $n_0$, namely $|\psi(0)\rangle=|n_0\rangle$, and then the probability at any site $m$ after evolution time $t$, $\rho_{m}(t)=|C_{m}(t)|^2$, takes~\cite{SM}, %$\rho_{m}(t)=|C_{m}(t)|^2$, takes~\cite{SM}
\begin{eqnarray}\label{rho}
\rho_{m}(t)=\mathcal{J}_{m-n_0}^{2}\left(2\sqrt{J_{L}J_{R}\left[u^{2}(t)+v^{2}(t)\right]}\right)\left(\frac{J_{R}}{J_{L}}\right)^{m-n_0},
\end{eqnarray}
with
\begin{eqnarray}\label{uv}
u(t)=\int_{0}^{t}dt'\cos\eta(t'),\quad v(t)=\int_{0}^{t}dt'\sin\eta(t').
\end{eqnarray}
%where $u(t)=\int_{0}^{t}dt'\cos\eta(t')$ and $v(t)=\int_{0}^{t}dt'\sin\eta(t').$
In the absence of electric fields, i.e., $E(t)=0$, from Eq. (\ref{eta}) and Eq. (\ref{uv}), we have $u(t)=t$ and $v(t)=0$, yielding $\rho_{m}(t)=\mathcal{J}_{m-n_0}^{2}\left(2t\sqrt{J_{L}J_{R}}\right)(J_{R}/J_{L})^{m-n_0}$, where $2t\sqrt{J_{L}J_{R}}$ linearly increases to infinity and thus $\mathcal{J}_{m-n_0}(2t\sqrt{J_{L}J_{R}})$ tends to $0$~\cite{Bsel} [see Fig.~\ref{01}(a)], such that $\rho_{m}/\rho_{m-1}\approx J_R/J_L$ when $t\rightarrow\infty$~\cite{SM}. Therefore, the system has right (left) boundary skin mode when $J_R/J_L>1$ ($J_R/J_L<1$), which is consistent with previous results~\cite{Yao2018,Kunst2018,Lee2019}. The following sections will discuss three cases: i) the pure dc field case, ii) the pure ac field case, and iii) the dc-ac mixed field case.

%%%%%%%%%%%%%%%%%%%%%%%%%%%%%%%%%%%%%%%%%%%%%%%%
%%\begin{widetext}
%%\end{widetext}
%%%%%%%%%%%%%%%%%%%%%%%%%%%%%%%%%%%%%%%%%%%%%%%%

{\em The pure dc electric field case.---} We firstly discuss the fate of NHSE in the presence of a pure dc electric field, i.e., $E(t)=E_{0}$. From Eq. (\ref{eta}) and Eq. (\ref{uv}), we have $u(t)=\sin (E_{0} t)/E_{0}$ and $v(t)=(1-\cos E_{0} t)/E_{0}$, and then, Eq. (\ref{rho}) gives the probability:
\begin{eqnarray}\label{dccase}
\rho_{m}(t)=\mathcal{J}_{m-n_0}^{2}\left(\frac{4\sqrt{J_{L}J_{R}}}{E_{0}}\sin\frac{E_{0} t}{2}\right)\left(\frac{J_{R}}{J_{L}}\right)^{m-n_0}.
\end{eqnarray}
Note that $\sin(E_{0} t^{*}/2)=0$ at the time points $t^{*}=2\pi N/E_{0}$ with $N=0,1,2,\cdots$. By using the properties of the Bessel function~\cite{Bsel} $\mathcal{J}_{m-n_0}(0)=\delta_{m-n_0,0}$ [see Fig.~\ref{01}(a)] and $(J_R/J_L)^{m-n_0}=1$ when $m=n_0$, we have that  $\rho_{m=n_0}(t^{*})$ will oscillate back to $1$ whatever the initial site $n_0$ is, and this phenomenon is called Stark localization~\cite{Wannier1962}, which induces that the particle initially localized at the bulk does not move toward the boundary. Therefore, the effect of the interplay between the Stark localization and NHSE is that even a small dc field is sufficient to suppress the NHSE.

The analytical results can be further confirmed by numerically calculating the winding number (WN). To define the WN, we need to introduce the twist boundary condition here, i.e., $\hat{H}(\Phi) = \hat{H} + J_Le^{i\Phi}|L\rangle\langle 1| + J_Re^{-i\Phi}|1\rangle\langle L|$, where $L$ is the system size and $\Phi$ is the introduced phase factor, and then the WN reads
\begin{equation}\label{winding}
  w=\frac{1}{2\pi i}\int_{0}^{2\pi}\partial_\Phi\ln\det\left[\hat{H}(\Phi)-\mathcal{E}_c\right]\ud\Phi.
\end{equation}
$w=1$ ($w=0$) corresponds to the existence (non-existence) of NHSE under OBC with eigenvalue around $\mathcal{E}_c$~\cite{Gong2018,Slager2020,Zensen,Okuma2020}, which is set to the algebra average of the spectrum in the following calculation. For convenience, we set $J_L=J-\gamma/2$ and $J_R=J+\gamma/2$ with $\gamma>0$ and $J=1$ as the unit of energy. Fig.~\ref{01} (b) shows the transition of the existence-nonexistence of NHSE, obtained by calculating the WN, which changes from $1$ to $0$ when the strength of dc field $E_0$ increases cross the transition line with fixed $L$ from left to right. It can be seen that the transition lines tend to $E_0=0$ with increasing $L$, which is consistent with the analytical result.

The analytical expression of $\rho_m(t)$ can also tell us the finite size effect, where interesting physics will emerge. Here, we consider $J_R>J_L$,
which makes the oscillation of the particle favor the right-hand side of the initial position $n_0$, and the oscillation range is approximate to $4J_R/|E_0|x_\star$~\cite{SM}, where $x_\star$ only depends on $J_L$ and $J_R$. If the distance between the right-side boundary and $n_0$ is larger than $4J_R/|E_0|x_\star$, the particle will return back to $n_0$ after a period of time, but if the distance is less than $4J_R/|E_0|x_\star$, the particle will arrive at the boundary and then stay there ever since. Thus, for fixed size $L$, there exist a critical electric field strength $E^c_0$ that describes the transition of the existence-nonexistence of NHSE and satisfies $4J_R/|E^c_0|x_\star=L$, giving $E_0^c\propto1/L$, as shown in the inset of Fig.~\ref{01} (b). When the size exceed a critical value, the number of skin modes is about $4J_R/|E_0|x_\star$, being independent of the size. It can be clearly seen by comparing Fig.~\ref{01} (c) and (d), which show the distributions of eigenstates with different sizes and same other parameters, and have the same number of skin modes.
Thus, with fixing $J_L$, $J_R$ and $E_0$, the number of skin modes can be of the same order of magnitude as the total eigenstate number for the system with small size and the NHSE exists.
When the size is large enough, the ratio of the number of skin modes to the total eigenstate number will be insignificant and the NHSE will disappear, suggesting that the NHSE is size-dependent, which is different from the general NHSE.
Moreover, we can increase the number of skin modes by decreasing the electric field strength, and thus, we can control the appearance or disappearance of NHSE for a finite size system, as shown in Fig.~\ref{01} (e) and (f), where the system shows NHSE when $E_0=0.005$, but when $E_0=0.5$, the NHSE disappear. The phenomenon of the size-dependent NHSE should widely exist in the non-Hermitian systems with defective, disordered, quasiperiodic or Stark potentials, or two coupled chains with dissimilar non-reciprocal hoppings~\cite{LinhuLi}.

\begin{figure}[t]
\hspace*{-0.1cm}
    \includegraphics[width=0.48\textwidth]{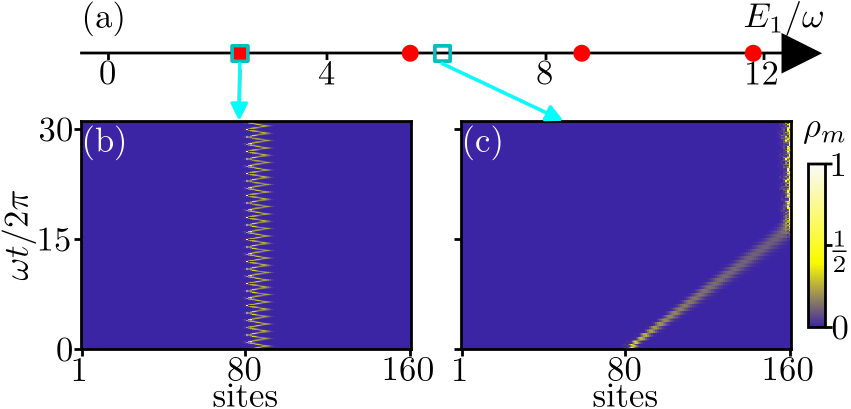}
    \caption{ (a) Red round dots are zeros of zeroth order of the Bessel function $\mathcal{J}_0(E_1/\omega)$, which correspond to the condition of the emergence of the dynamic localization under ac field driving.
    Dynamical evolution of a particle initially localized at the lattice center with
    (b) $E_1/\omega=2.405$, which is the first zero point of $\mathcal{J}_0(E_1/\omega)$ and (c)
    $E_1/\omega=6.1$.  Here we set $J=1$, $\gamma=0.73$, and $\omega=0.46$.}
    \label{02}
\end{figure}

{\em The pure ac electric field case.}--- We then consider the monochromatic
cosine shape ac electric field $E(t)=E_1\cos(\omega t)$. It is a typical Floquet driving system, with period $T=2\pi/\omega$. From Eq. (\ref{eta}) and Eq. (\ref{uv}), we have $u(t)=\int_{0}^{t}dt'\cos\left(\frac{E_{1}}{\omega}\sin\omega t'\right)$ and $v(t)=\int_{0}^{t}dt'\sin\left(\frac{E_{1}}{\omega}\sin\omega t'\right)$, and then, Eq. (\ref{rho}) gives the probability
in the limit $t\gg T$~\cite{SM},
\begin{eqnarray}
    \rho_{m}(t\gg T)
    \approx\mathcal{J}_{m-n_0}^{2}\left(2t\sqrt{J_{L}J_{R}}
    \mathcal{J}_{0}\left(\frac{E_{1}}{\omega}\right)\right)\left(\frac{J_{R}}{J_{L}}\right)^{m-n_0}.&&\nonumber\\
    &&
    \label{solac}
\end{eqnarray}
%{\color{blue} Because $\mathcal{J}_0(x)$ is an even function, our result below is not dependent on the direction of the ac field.}
When $\mathcal{J}_{0}({E_{1}}/{\omega})\neq 0$,
$2t\sqrt{J_{L}J_{R}}\mathcal{J}_{0}({E_{1}}/{\omega})$ increases linearly to infinity and thus $\mathcal{J}_{m-n_0}(2t\sqrt{J_{L}J_{R}}\mathcal{J}_{0}({E_{1}}/{\omega}))$ tends to zero, which is completely similar to the case without electric field, suggesting that the NHSE is not affected by the ac field. For the special ac field strength $E_1$ and frequency $\omega$ that satisfy $\mathcal{J}_{0}({E_{1}}/{\omega})= 0$, corresponding to the red round dots in Fig.~\ref{02}(a),
due to $\mathcal{J}_{m-n_0}(0)=\delta_{m-n_0,0}$, the particle initially localized in the bulk will move around the initial position, and thus, the NHSE will be suppressed, as shown in Fig.~\ref{02}(b). This localization phenomenon is called dynamic localization~\cite{Dunlap1986,ZhaoXG,Longhi2006,Eckardt2009}, which is distinct from Anderson localization induced by random disorder potential. For most cases, $\mathcal{J}_{0}({E_{1}}/{\omega})\neq 0$,
the particle will hop to the boundary eventually as demonstrated in Fig.~\ref{02}(c). To sum up, when only applying the ac electric field to the system, the NHSE is not affected except for these special parameters of $E_1$ and $\omega$ that causes the dynamic localization, suppressing the NHSE.

\begin{figure}[t]
\hspace*{-0.1cm}
    \includegraphics[width=0.48\textwidth]{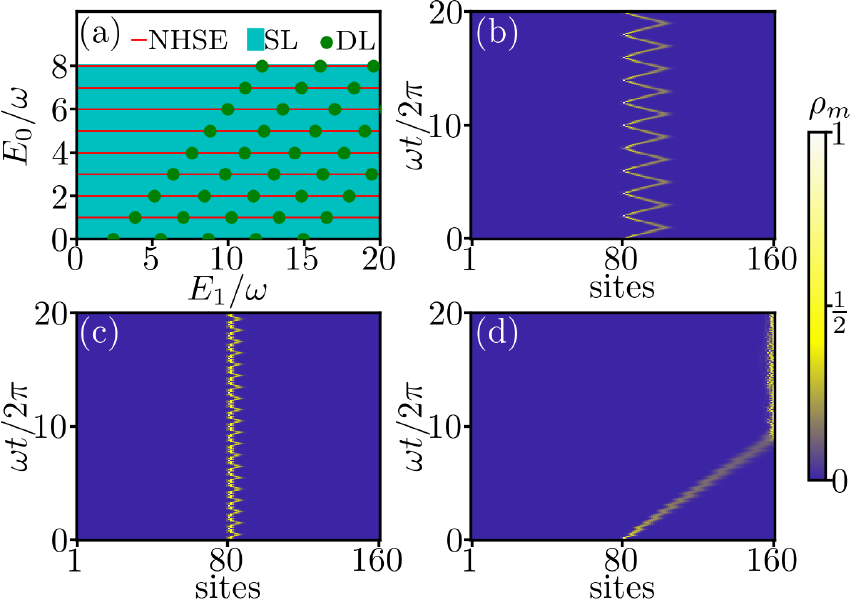}
    \caption{(a) The schematic phase diagram about the existence or non-existence of NHSE under
    the mixed electric field driving. Red lines represent the integer $E_0/\omega$. Dark green dots on the red lines correspond to $E_1/\omega$ being the zeroes of $\mathcal{J}_{E_0/\omega}$, where dynamic localization (DL) induced by the ac field occurs.  The light green region correspond to the Stark localization (SL).
    The dynamical evolutions of the electron initialized on the lattice center under simultaneous driving of dc and ac fields with
    (b) $E_0/\omega=0.5$ and $E_1/\omega=1.3$, (c) $E_0/\omega=1$  and $E_1/\omega=3.832$ corresponding to the first zero of
    $\mathcal{J}_1$, and (d) $E_0/\omega=1$ and $E_1/\omega=5.7$.
    Here we fix $J=1$, $\gamma=0.73$ and $\omega=0.46$.}
    \label{03}
\end{figure}

{\em The dc+ac electric field case.}--- We finally consider the effect of dc-ac mixed fields, i.e., $E(t)=E_0+E_1\cos(\omega t)$, in manipulating NHSE. By using Eq. (\ref{eta}) and Eq. (\ref{uv}), we can calculate functions $u(t)$ and $v(t)$, whose expressions look fair complicated~\cite{SM}, and find that when $E_{0}/\omega$ is not an integer, all the terms in the expressions of  $u(t)$ and $v(t)$ are bounded oscillatory functions of time, meaning that the particle will oscillate around the initial position. Therefore, the electric fields break NHSE for non-integer $E_{0}/\omega$.
When $E_{0}/\omega$ is an integer, the probability
in the limit of long evolution time~\cite{SM} can be simplified to
\begin{equation}
    \rho_{m}(t\gg T)
    \approx
    \mathcal{J}_{m-n_0}^{2}\left(2t\sqrt{J_{L}J_{R}}\mathcal{J}_{\frac{E_{0}}{\omega}}\left(\frac{E_{1}}{\omega}\right)\right)\left(\frac{J_{R}}{J_{L}}\right)^{m-n_0}.
    \label{soldcac}
\end{equation}
Similar to the discussions for Eq. (\ref{solac}), the disappearance or existence of NHSE depends on whether $E_1/\omega$ is one of the zeroes of the Bessel function $\mathcal{J}_{E_0/\omega}$. The pure dc field can cause the Stark localization, which suppresses NHSE. Then adding the ac field $E_1\cos(\omega t)$ with $E_0/\omega$ being integers, the particle can break though the localization barrier and move through the chain accompanied by the photon absorption or emission. Thus, NHSE can exist only for integer $E_0/\omega$ but except the situations that $E_1/\omega$ are the zeroes of $\mathcal{J}_{E_0/\omega}$, which will induce the dynamic localization, as discussed above. The effects of the dc-ac mixture fields on NHSE are summarized in the Fig.~\ref{03}(a). The light green region between the red lines and the dark green dots on the red lines respectively correspond to the non-integer $E_0/\omega$ and the zeroes of $\mathcal{J}_{E_0/\omega}$ with $E_0/\omega$ being integers, which will induce the Stark localization and dynamic localization and lead to the disappearance of NHSE, as shown in Fig.~\ref{03}(b) and (c). The red lines correspond to the integer $E_0/\omega$, which can break the bulk localization by photon assisted hopping, and the particle initially localized in the bulk will move toward the boundary, as shown in Fig.~\ref{03}(d).

\begin{figure}[t]
\hspace*{-0.1cm}
  %  \begin{center}
        \includegraphics[width=0.47\textwidth]{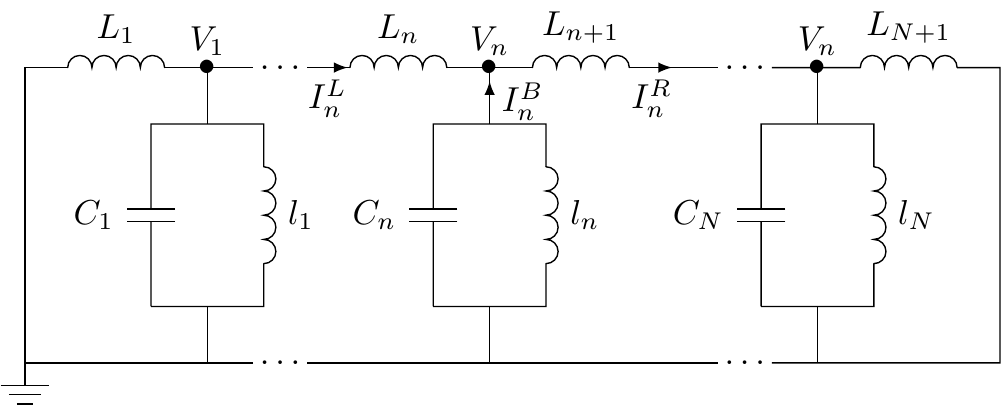}
        \caption{Schematic of the LC electronic circuit.}
        \label{04}
  %  \end{center}
\end{figure}

{\em Electronic circuit's realization.---}The non-Hermitian model (\ref{hams}) can be simulated by a
classical electric circuit as depicted in Fig.~\ref{04}, which consists of $L$ LC circuit units.
Based on the Kirchhoff's current law, we have
\begin{equation}
    I_n^R = I_n^L + I_n^B,
    \label{law}
\end{equation}
where $I_n^B$ is the current flows through the $n$th unit and $I_n^L$ ($I_n^R$) is the current
from the $(n-1)$th ($n$th) unit to the $n$th ($(n+1)$th) unit, and they satisfy
\begin{eqnarray}
    L_n\frac{\ud{}I_n^{L}}{\ud{}t} = V_n-V_{n-1},
    \quad
    L_{n+1}\frac{\ud{}I_n^{R}}{\ud{}t} = V_{n+1}-V_{n},\label{I1}\\
    l_n\frac{\ud}{\ud{}t}\left[I_n^B - C_n\frac{\ud{}V_n}{\ud{}t}\right]
    =V_n,\qquad \qquad
    \label{I2}
\end{eqnarray}
where $V_n$ is the voltage on the node $n$. From Eqs. (\ref{law}-\ref{I2}),
one can obtain,
\begin{equation}
    \frac{\ud^2V_n}{\ud{}t^2}
    =\frac{V_{n+1}-V_n}{C_nL_{n+1}}+\frac{V_{n-1}-V_n}{C_nL_{n}}
    -\frac{V_n}{C_nl_n}
    \label{Vn}
\end{equation}

When choosing inductors with inductances $L_n=L_0g^{-n}$, $l_n=L_n/(\Delta-anE_0)$, and capacitors with capacitance $C_n=C_0g^n$, Eq. (\ref{Vn}) becomes
\begin{equation}
    \left(1+g+\Delta+\frac{1}{\omega_0^2}\frac{\ud^2}{\ud{}t^2}\right)V_n
    =V_{n-1}+gV_{n+1}+naE_0V_n,
    \label{eqHam}
\end{equation}
where $\omega_0=1/\sqrt{C_0L_0}$. We make a transformation: $V_n\rightarrow V_ne^{\pm{}i\omega{}t}$ with $n=1,2,\dots,L$ and
$\omega = \omega_0\sqrt{(1+g+\Delta)-E_R}$, then Eq. (\ref{eqHam}) becomes $E_RV_n=V_{n-1}+gV_{n+1}+naE_0V_n$,
which describes the dc case of the model (\ref{hams}) with $J_L=1$ and $J_R=g$. By directly detecting the eigenvalues and eigenstates through an elementary voltage measurement~\cite{Helbig2020}, one can detect the manipulation effect of dc field.

We note that our results can also apply to the manipulation of the NHSE induced by the on-site dissipations.
Recent experiment~\cite{Yi2021} and theoretical proposals~\cite{Yangzs,Zhoulh} suggest realizing and detecting the NHSE in the dissipative ultracold atom systems, where the gradient fields can be easily realized~\cite{Arimondo2008,Wolfgang2013,Ibloch}. Therefore, the control effect of the electric fields on the NHSE can be detected in optical lattices. Furthermore, this control effect can also be detected based on the photonic quantum walk~\cite{Xiao2020} and the sideband cooling setups in trapped ion systems~\cite{ionRMP,ionWang}.

{\em Summary and discussion.---} We have investigated the control effect of electric fields on NHSE analytically and numerically. For the pure dc field case, in the thermodynamic limit, a weak Stark localization induced by the dc field is sufficient to win the competition with NHSE, so a non-zero dc field can suppress the NHSE. When the system size is finite, the new interesting phenomenon of the size-dependent NHSE will emerge, because the number of skin modes is size-independent when the size exceed a critical value. For the pure ac field case, only the special field strength $E_1$ and frequency $\omega$ that satisfy $\mathcal{J}_{0}({E_{1}}/{\omega})= 0$ can suppress the NHSE due to the dynamic localization. For the mixed field case, if $E_0/\omega$ is not an integer, the NHSE will be suppressed by the Stark localization induced by the dc field. For the integer $E_0/\omega$, NHSE can exist, because the particle can move toward the boundary by the photon absorption or emission except for the special case with $E_1/\omega$ being one of the zeroes of $\mathcal{J}_{E_0/\omega}$, which causes the dynamic localization.

The control effects of electric fields on the NHSE are abundant, and moreover, electric fields can be easily applied to a system. Thus, the manipulation methods can be widely used in experiments and the fabrication of new devices. For instance, based on the phenomenon of the size-dependent NHSE and the sensitivity of the NHSE versus the field strength and frequency of ac fields near the special situations that satisfy $\mathcal{J}_{0}({E_{1}}/{\omega})= 0$, by detecting the signals on the boundary of a non-Hermitian system, one can carry out accurate measurements of electric fields, which are important for many critical applications in science and industry. As a second example, the NHSE can be used to design some devices, such as directional amplifiers~\cite{Metelmann,Metelmann2,Peterson2017,Barzanjeh,McDonald2018,Malz2018,Harris2019,Wentan2021} and light funnels~\cite{Weidemann2020}. The appearance or disappearance of the directional amplification and the funneling effect can be manipulated by using dc or ac fields, such as by changing the strength or frequency of an added ac field. Thus, one can design the switch of these devices by using electric fields.

\begin{acknowledgments}
We thank L. Li and H. Jiang for valuable discussions. This work was supported by the National Natural Science Foundation of China (Grants No. U1801661, No.12104205, No.12104210), the Key-Area Research and Development Program of Guangdong Province (Grant No. 2018B030326001) ,Guangdong Provincial Key Laboratory (Grant No.2019B121203002).
\end{acknowledgments}
%the Science, Technology and Innovation Commission of Shenzhen Municipality (No.KYTDPT20181011104202253) ,Grant No.2016ZT06D348,
%the Natural Science Foundation of Guangdong Province (Grant No. 2017B030308003)
%%%%%%%%%%%%%%%%%%%%%%%%%%%%%%%%%%%%%%%%%%%%%

%%%%%%%%%%%%%%%%%%%%%%%%%%%%%%%%%%%%%%%%%%%%%%%%%%%%%
%%%%%%%%%%%%%%%%%%%%%%%%%%%%%%%%%%%%%%%%%%%%%%%%%%%%%%
\global\long\def\id{\mathbbm{1}}
\global\long\def\ui{\mathbbm{i}}
\global\long\def\ud{\mathrm{d}}

%%%%%%%%%%%%%%%%%%%%%%%%%%%%%%%%%%%%%%%%%%%%%%%%%%%%%%%%%%%%%%%
\setcounter{equation}{0} \setcounter{figure}{0}
\setcounter{table}{0} %\setcounter{page}{1} \makeatletter
\renewcommand{\theparagraph}{\bf}
\renewcommand{\thefigure}{S\arabic{figure}}
\renewcommand{\theequation}{S\arabic{equation}}

\onecolumngrid
\flushbottom
%%%%%%%%%%%%%%%%%%%%%%%%%%%%%%%%%%%%%%%%%%%%%%%%
\newpage
\section*{\large Supplementary Material:\\Manipulating non-Hermitian skin effect via electric field}
%\section*{\normalsize SUPPLEMENTAL MATERIAL}
In the Supplementary Materials, we first give the details of deriving the analytical solutions of quantum dynamics. Then, we discuss the analytical results and obtain the effects of the pure dc field, pure ac field and the mixed field on the non-Hermitian skin effect. Finally, we discuss the finite size effect in the pure dc field case and the dynamical behavior in the absence of external fields.

\section{I. Analytical solutions of quantum dynamics}

We derive the analytical solutions of quantum dynamics for our one-dimensional non-Hermitian model driven by an arbitrary time-dependent electric field $E(t)$. The Hamitonian are following,
\begin{eqnarray}
\hat{H}(t)=\sum_{m=-\infty}^{+\infty}\left(J_{L}|m\rangle\langle m+1|+J_{R}|m+1\rangle\langle m|\right)+E(t)\sum_{m=-\infty}^{+\infty}m|m\rangle\langle m|,\label{hami}
\end{eqnarray}
where $J_{R}$ and $J_{L}$ are the strengths of the leftforward and rightforward hopping, respectively. %$f(t)$ is an arbitrary time-dependent electric field.
We expand the time-dependent quantum state as $|\psi(t)\rangle=\sum_{m=-\infty}^{+\infty}C_{m}(t)|m\rangle$ and  substitute it into Schr\"{o}dinger equation $i \partial_t |\psi(t)\rangle =\hat{H}(t)|\psi(t)\rangle$ (we set $\hbar=1$ throughout this Supplementary Material) to arrive at the equation of the time-dependent amplitudes $C_{m}(t)=\langle m|\psi(t)\rangle$,
\begin{eqnarray}
i \partial_t C_{m}(t) = J_{L}C_{m+1}(t)+J_{R}C_{m-1}(t)+E(t)mC_{m}(t). \label{eqcm}
\end{eqnarray}
In solving Eq. (\ref{eqcm}), we transfer to the momentum space firstly by the discrete Fourier transformation,
\begin{eqnarray}
C_{k}(t)=\sum_{m=-\infty}^{+\infty}e^{-ikm}C_{m}(t),
\end{eqnarray}
and then we can directly rewrite Eq. (\ref{eqcm}) as
\begin{eqnarray}
\partial_t C_{k}(t)-E(t)\partial_k C_{k}(t)=-i\left(J_{L}e^{ik}+J_{R}e^{-ik}\right)C_{k}(t).\label{eqck}
\end{eqnarray}
This partial differential equation can be transferred to an ordinary differential equation by the variables transformation
\begin{eqnarray}
p=t, q=k+\eta(t),
\end{eqnarray}
with $\eta(t)=\int_{0}^{t}E(t')dt'$. This gives rise to  $C_{k}(t)=C(p,\tau)$ with $\tau=q-\eta(p)$ and the chain rule of the derivation of $C(p,\tau)$ over $p$ results in the ordinary differential equation,
\begin{eqnarray}
%\frac{d C_k(t)}{dt}&=&
\frac{d C(p,\tau)}{dp}&=&\frac{\partial C(p,\tau)}{\partial p}\frac{\partial p}{\partial p}+\frac{\partial C(p,\tau)}{\partial \tau}\frac{\partial \tau}{\partial p},\nonumber\\
&=&\frac{\partial C(p,\tau)}{\partial p}-E(p)\frac{\partial C(p,\tau)}{\partial \tau},\nonumber\\
&=&-i\left(J_{L}e^{i\tau}+J_{R}e^{-i\tau}\right)C(p,\tau), \label{eqpq}
\end{eqnarray}
where we applied Eq. (\ref{eqck}) to the last step. Eq. (\ref{eqpq}) can be solved by integrating $p$ over two sides and then
\begin{eqnarray}
C(p,\tau)&=&C(0,\tau|_{p=0})e^{-i\int_{0}^{p}\left[J_{L}e^{i\left(q-\eta(p')\right)}+J_{R}e^{-i\left(q-\eta(p')\right)}\right]dp'},\nonumber\\
&=&C(0,\tau|_{p=0})e^{-i\int_{0}^{p}\left[(J_{L}+J_{R})\cos\left(q-\eta(p')\right) +i (J_{L}-J_{R})\sin\left(q-\eta(p')\right) \right]dp'},\nonumber\\
&=&C(0,q)e^{-i\int_{0}^{p}\left[ J_{+}\cos (q-\eta(p'))+iJ_{-}\sin (q-\eta(p'))\right]dp'},\label{cpt}
\end{eqnarray}
where $J_{\pm}=J_{L}\pm J_{R}$. Transferring Eq. (\ref{cpt}) back to the form with variables $(k,t)$ yields the solution of Eq. (\ref{eqck}),
\begin{eqnarray}
C_{k}(t)=C_{k+\eta(t)}(0)e^{-i\int_{0}^{t}\left[J_{+}\cos(k+\eta(t)-\eta(t'))+iJ_{-}\sin(k+\eta(t)-\eta(t'))\right]dt'}.\label{solck}
\end{eqnarray}
To simplify the solutions, we define the following functions
\begin{eqnarray}
\mathcal{U}(t)&=&\int_{0}^{t}\cos\left[\eta(t)-\eta(t')\right]dt',\\
\mathcal{V}(t)&=&\int_{0}^{t}\sin\left[\eta(t)-\eta(t')\right]dt',
\end{eqnarray}
and the Eq. (\ref{solck}) arrives at
\begin{eqnarray}
C_{k}(t)&=&C_{k+\eta(t)}(0)e^{-i\left[J_{+}(\mathcal{U}\cos k-\mathcal{V}\sin k)+iJ_{-}(\mathcal{U}\sin k+\mathcal{V}\cos k))\right]dt'},\nonumber\\
&=&C_{k+\eta(t)}(0)e^{i\left(\overline{\mathcal{V}}\sin k-\overline{\mathcal{U}}\cos k\right)}.\label{solck1}
\end{eqnarray}
where
\begin{eqnarray}
\overline{\mathcal{U}}(t)=J_{+}\mathcal{U}(t)+i J_{-}\mathcal{V}(t), \overline{\mathcal{V}}(t)=J_{+}\mathcal{V}(t)-i J_{-}\mathcal{U}(t). \label{uvde1}
\end{eqnarray}
We apply the following expansions, which is expanded by ordinary Bessel functions $\mathcal{J}_{m}(x)$, to Eq. (\ref{solck1})  \cite{BselS},
\begin{eqnarray}
e^{-i\overline{\mathcal{U}}\cos k}&=&\sum_{m=-\infty}^{+\infty}e^{-i\frac{m\pi}{2}}e^{imk}\mathcal{J}_{m}(\overline{\mathcal{U}}),\\
e^{i\overline{\mathcal{V}}\sin k}&=&\sum_{m=-\infty}^{+\infty}e^{imk}\mathcal{J}_{m}(\overline{\mathcal{V}}),
\end{eqnarray}
and transfer the solutions in Eq. (\ref{solck1}) back to the spatial space by the  discrete Fourier transformation,
\begin{eqnarray}
C_{m}(t)=\sum_{k}e^{ikm}C_{k}(t).
\end{eqnarray}
Thus the time-dependent amplitude in spatial space $C_{m}(t)$ are
\begin{eqnarray}
C_{m}(t)&=&\sum_{k}e^{ikm}C_{k+\eta(t)}(0)e^{i\left(\overline{\mathcal{V}}\sin k-\overline{\mathcal{U}}\cos k\right)},\nonumber\\
&=&\sum_{k}\sum_{n,r=-\infty}^{+\infty}e^{ik(m+n+r)}C_{k+\eta(t)}(0)e^{-i\frac{n\pi}{2}}J_{n}(\overline{\mathcal{U}})J_{r}(\overline{\mathcal{V}}),\nonumber\\
&=&\sum_{k}\sum_{n,r=-\infty}^{+\infty}e^{i(k+\eta(t))r}C_{k+\eta(t)}(0)e^{-i\eta(t)r}e^{-i\frac{n\pi}{2}}J_{n}(\overline{\mathcal{U}})J_{r-n-m}(\overline{\mathcal{V}}),\nonumber\\
&=&\sum_{n,r=-\infty}^{+\infty}C_{r}(0)e^{-i\eta(t)r}e^{-i\frac{n\pi}{2}}J_{n}(\overline{\mathcal{U}})J_{r-n-m}(\overline{\mathcal{V}}),\nonumber\\
&=&\sum_{n,r=-\infty}^{+\infty}C_{r}(0)e^{-i\eta(t)r}e^{-i\frac{n\pi}{2}}J_{n}(\overline{\mathcal{U}})e^{-i(n+m-r)\pi}J_{n+m-r}(\overline{\mathcal{V}}),\nonumber\\
&=&\sum_{r=-\infty}^{+\infty}(-1)^{m-r}C_{r}(0)e^{-i\eta(t)r}\left[ \sum_{n=-\infty}^{+\infty} J_{n}(\overline{\mathcal{U}})J_{n+m-r}(\overline{\mathcal{V}})e^{i\frac{n\pi}{2}}\right],\nonumber\\
&=&\sum_{r=-\infty}^{+\infty}(-1)^{m-r}C_{r}(0)e^{-i\eta(t)r}\left(\frac{\overline{\mathcal{V}}+i \overline{\mathcal{U}}}{\overline{\mathcal{V}}-i \overline{\mathcal{U}}}\right)^{\frac{m-r}{2}}J_{m-r}\left(\sqrt{\overline{\mathcal{U}}^2+\overline{\mathcal{V}}^2}\right).\label{fsol0}
\end{eqnarray}
In the last step, we applied the Graf's addition theorem of the Bessel functions \cite{BselS}.  Through the relations in Eq. (\ref{uvde1}), we can straightforwardly obtain that
\begin{eqnarray}
\frac{\overline{\mathcal{V}}+i \overline{\mathcal{U}}}{\overline{\mathcal{V}}-i \overline{\mathcal{U}}}=\frac{J_{R}}{J_{L}}\frac{i \mathcal{V}-\mathcal{U}}{i \mathcal{V}+\mathcal{U}},~~~~~~\overline{\mathcal{U}}^2+\overline{\mathcal{V}}^2=2\sqrt{J_{L}J_{R}(\mathcal{U}^{2}+\mathcal{V}^{2})}=2\sqrt{J_{L}J_{R}(u^{2}+v^{2})}.
\end{eqnarray}
Those expressions simplify the solution of Eq. (\ref{fsol0}) to
\begin{eqnarray}
C_{m}(t)&=&\sum_{n=-\infty}^{+\infty}(-1)^{m-n}C_{n}(0)e^{-i\eta(t)n}J_{m-n}\left(2\sqrt{J_{L}J_{R}\left[\mathcal{U}^{2}(t)+\mathcal{V}^{2}(t)\right]}\right)\left[\frac{J_{R}}{J_{L}}\frac{i \mathcal{V}(t)-\mathcal{U}(t)}{i \mathcal{V}(t)+\mathcal{U}(t)}\right]^{\frac{m-n}{2}}.\label{fsol}
\end{eqnarray}
This solution is valid for arbitrary initial state and for any time-dependent driven $f(t)$, and also can be specified to the system with finite long chains.
\section{II. Discussions of the analytical results}
Base on the exact analytical solution of quantum evolution in Eq. (\ref{fsol}), we can study the properties of the skin-effect and the electric fields induced Wannier Stark localization by choosing an appropriate initial state. Here, we set the system in site $m=0$ as the specific initial state, i.e., $C_{m=0}(t=0)=1$, this is equivalent to set any other sites $m\neq 0$ as the initial state for the infinity long chain. Therefore, we can simplify the solution in Eq. (\ref{fsol}) and explicitly express the probability as,
\begin{eqnarray}
\rho_{m}(t)=|C_{m}(t)|^2=\mathcal{J}_{m}^{2}\left(2\sqrt{J_{L}J_{R}\left[u^{2}(t)+v^{2}(t)\right]}\right)\left(\frac{J_{R}}{J_{L}}\right)^m,\label{fsol1}
\end{eqnarray}
where we used the relation $\mathcal{U}^{2}+\mathcal{V}^{2}=u^{2}+v^{2}$ and here
\begin{eqnarray}
u(t)=\int_{0}^{t}dt'\cos\eta(t'),~~~~~ v(t)=\int_{0}^{t}dt'\sin\eta(t').
\end{eqnarray}
We note that Eq. (\ref{fsol1}) will become Eq.(5) of the main text when we choose the initial state localized at $n_0$, i.e., $C_{m=n_0}(t=0)=1$
instead of $C_{m=0}(t=0)=1$. The exact solutions in Eq. (\ref{fsol1}) are valid when $J_{L}J_{R}\neq0$ and project to Hermitian case when $J_{L}=J_{R}$. In the Non-Hermitian cases $J_{L}\neq J_{R}$, the coefficients $(J_{R}/J_{L})^m$ are responsible to the skin effect. Eq. (\ref{fsol1}) is normalized for the Hermitian case, and for non-Hermitian cases, the addition normalization is needed.

\begin{figure}[t]
\centering \includegraphics[width=18cm]{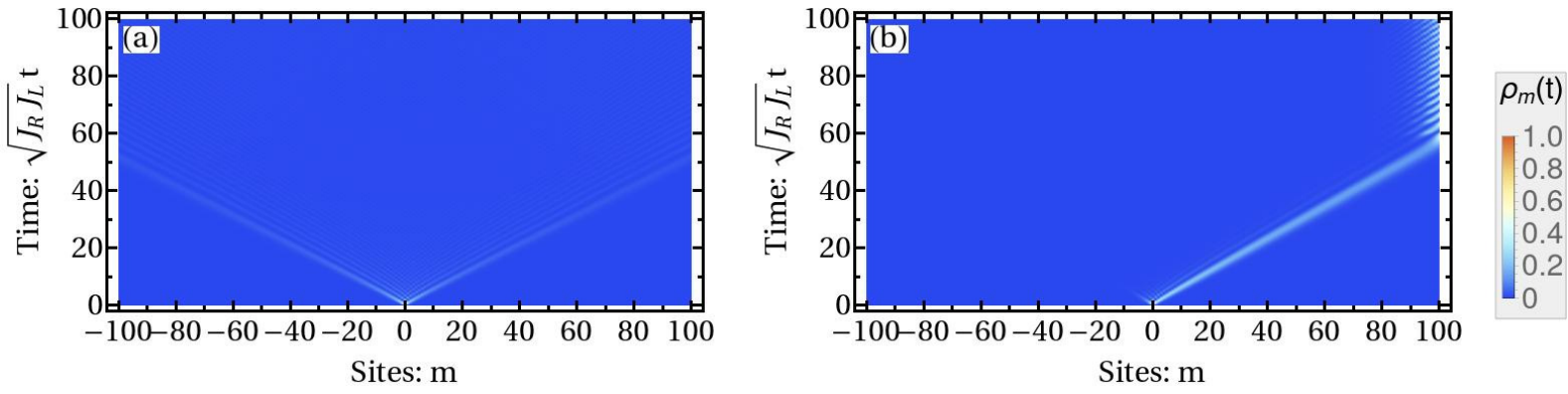}
\caption{ The quantum dynamics in the absence of fields for Hermitian case$~(a):J_{L}=J_{R}=1.0$, and for the Non-Hermitian case$~(b):J_{L}=0.8, J_{R}=1.0$. }
\label{figS1}
\end{figure}

Next, we consider the following electric fields
\begin{eqnarray}
E(t)=E_0+E_1 \cos\omega t,
\end{eqnarray}
where $E_{0}$ and $E_{1}$ are the strengths for the dc and ac parts, respectively. $\omega$ is the ac driven frequency. Under this driven, we have
\begin{eqnarray}
\eta(t)=E_{0}t+\frac{E_{1}}{\omega}\sin\omega t,
\end{eqnarray}
and then
\begin{eqnarray}
u(t)=\int_{0}^{t}dt'\cos\left(E_{0}t'+\frac{E_{1}}{\omega}\sin\omega t'\right),~~~~~ v(t)=\int_{0}^{t}dt'\sin\left(E_{0}t'+\frac{E_{1}}{\omega}\sin\omega t'\right).\label{genuv}
\end{eqnarray}
$u(t)$ and $v(t)$ are not bounded functions only in the absence of electric fields, namely $E(t)=0$.
%and hopping terms exist in the Hamitonian (\ref{hami}),
 In this case, we have $u(t)=t$ and $v(t)=0$, yielding $\rho_{m}(t)=\mathcal{J}_{m}^{2}\left(2t\sqrt{J_{L}J_{R}}\right)(J_{R}/J_{L})^m$, which gives purely skin effect due to the argument $x$ in $\mathcal{J}_{m}(x)$ is linearly increasing to infinity and then $\mathcal{J}_{m}(x)$ approaches to vanish, such that only the probabilities of the edge sites survive by the coefficients $(J_{R}/J_{L})^m$. This purely skin effect is shown in Fig. \ref{figS1}(b) through the quantum dynamic and as a comparison, the Hermitian case is shown in Fig. \ref{figS1}(a).
\subsection{DC electric field}
\begin{figure*}[t]
\centering \includegraphics[width=18cm]{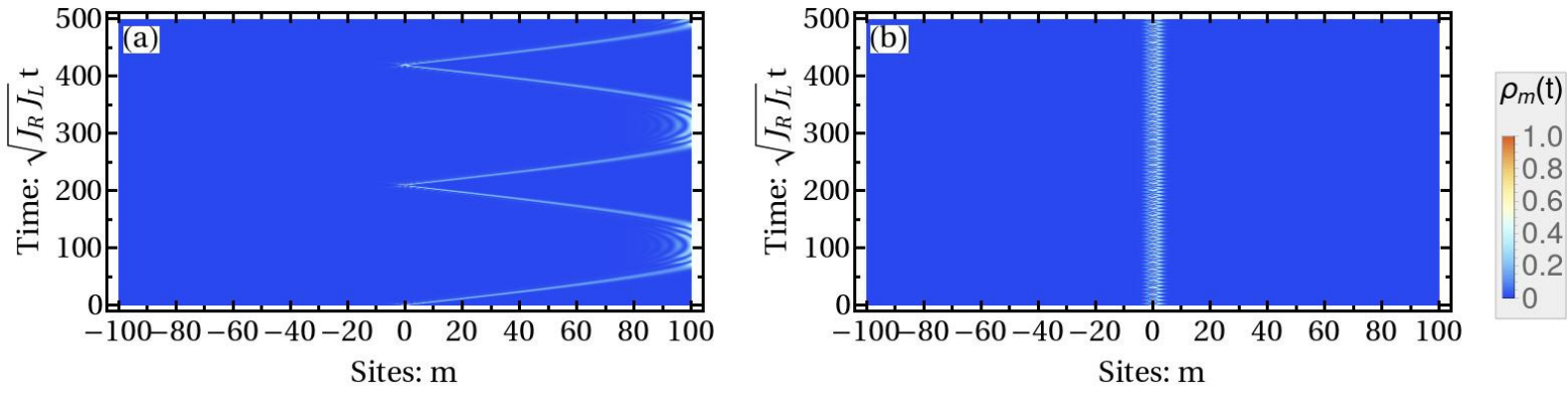}
\caption{ The quantum dynamics for the dc fields with the weakly driven$~(a):E_{0}=0.03$, and the strongly driven$~(b):E_{0}=1.0$. Other parameters are $J_{L}=0.8, ~J_{R}=1.0,~ \omega=1.0$.}
\label{figS2}
\end{figure*}

For this time-independent driven case $E(t)=E_{0}$, we have
\begin{eqnarray}
u(t)=\frac{\sin E_{0} t}{E_{0}},~~~~~ v(t)=\frac{1-\cos E_{0} t}{E_{0}},
\end{eqnarray}
and the probability is
\begin{eqnarray}
\rho_{m}(t)=\mathcal{J}_{m}^{2}\left(\frac{4\sqrt{J_{L}J_{R}}}{E_{0}}\sin\frac{E_{0} t}{2}\right)\left(\frac{J_{R}}{J_{L}}\right)^m.\label{dcsol}
\end{eqnarray}
From Eq. (\ref{dcsol}), the argument of $\mathcal{J}_{m}(x)$ is bounded and oscillating along time when $E_{0}\neq 0$. It means that the probability amplitude $\rho_{0}$ will oscillate back to 1 at the time points $t^{*}=2\pi N/E_{0}$ with $N=0,1,2,\cdots$, due to the fact that $\mathcal{J}_{m}(0)=\delta_{m0}$. This Wannier Stark localization will compete with skin effect under the weakly driven shown in Fig. \ref{figS2}(a), and dominate the dynamics for the strongly driven shown in Fig. \ref{figS2}(b).
\subsection{AC electric field}
\begin{figure*}[t]
\centering \includegraphics[width=18cm]{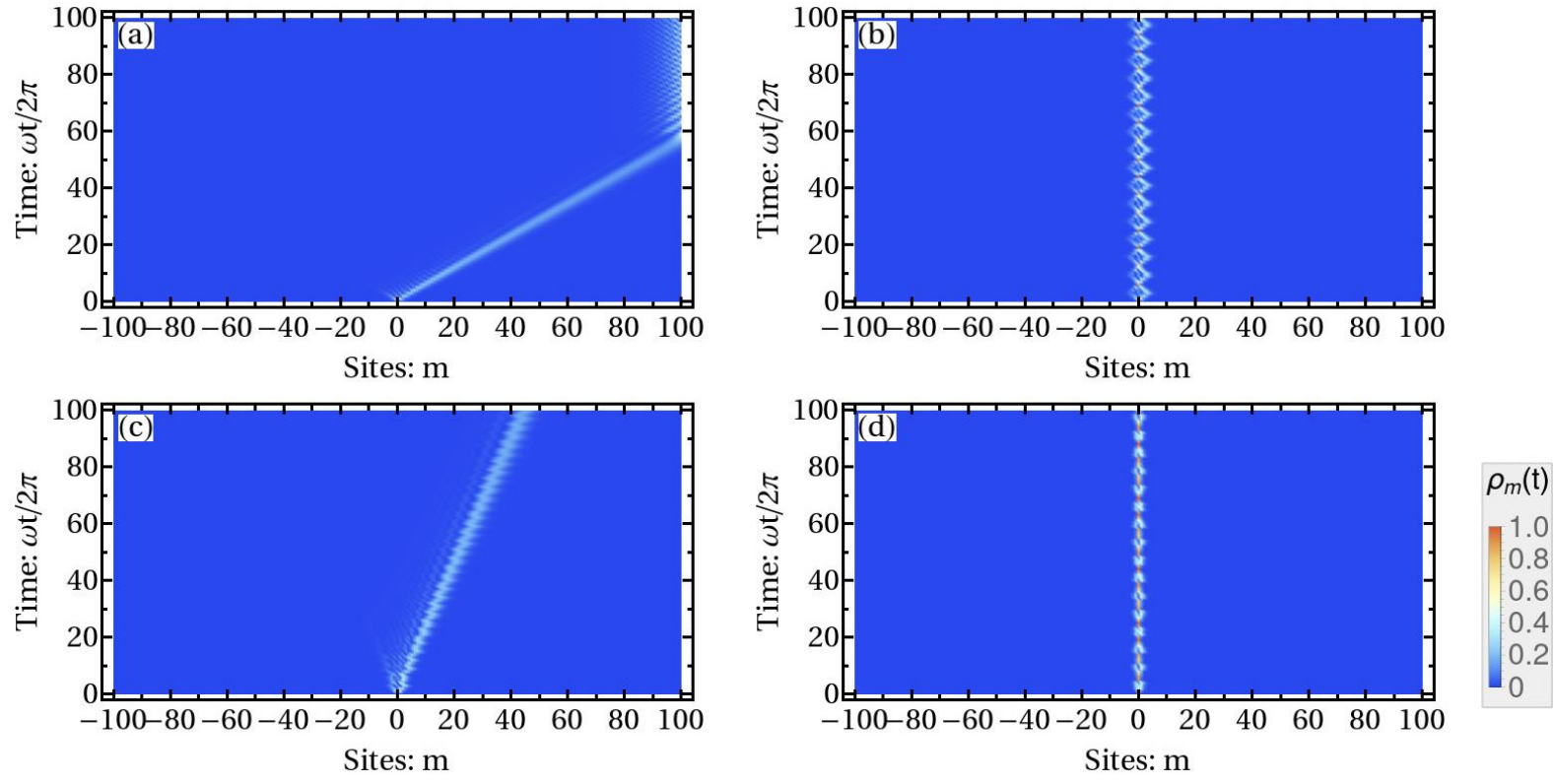}
\caption{The quantum dynamics for the ac fields with the ratios$~(a):E_{1}/\omega=0.1,~(b):0.2402~ (\text{1st zero of~} \mathcal{J}_{0}),~(c):3.0,~(d):5.52~ (\text{2nd zero of~} \mathcal{J}_{0}(x))$. Other parameters are $J_{L}=0.8, ~J_{R}=1.0,~ \omega=1.0$.}
\label{figS3}
\end{figure*}

For this time-dependent driven case $E(t)=E_{1}\cos\omega t$, we have
\begin{eqnarray}
u(t)=\int_{0}^{t}dt'\cos\left(\frac{E_{1}}{\omega}\sin\omega t'\right),~~~~~ v(t)=\int_{0}^{t}dt'\sin\left(\frac{E_{1}}{\omega}\sin\omega t'\right).
\end{eqnarray}
At each periodic cycles, i.e., $t^{*}=2\pi N/\omega$ with $N=0,1,2,\cdots$, we have
\begin{eqnarray}
u(t^{*})=\frac{t^{*}}{\pi}\int_{0}^{\pi}d\tau\cos\left(\frac{E_{1}}{\omega}\sin\tau\right)=t^{*}\mathcal{J}_{0}(\frac{E_{1}}{\omega}),~~~~~ v(t)=0,
\end{eqnarray}
and such that the probability at time $t^{*}$ reads
\begin{eqnarray}
\rho_{m}(t^{*})=\mathcal{J}_{m}^{2}\left(2t^{*}\sqrt{J_{L}J_{R}}\mathcal{J}_{0}(\frac{E_{1}}{\omega})\right)\left(\frac{J_{R}}{J_{L}}\right)^m,
\end{eqnarray}
where shows that the system is localized only when $E_{1}/\omega$ is the zeros of $J_{0}$, otherwise the skin effect will dominate the dynamics in the long time. To explicitly present this, we define two bounded functions as
\begin{eqnarray}
\overline{u}(t)=u(t)-t\mathcal{J}_{0}(\frac{E_{1}}{\omega}),~~~\overline{v}(t)=v(t)-0.
\end{eqnarray}
Then we rewrite the probability as
\begin{eqnarray}
\rho_{m}(t)&=&\mathcal{J}_{m}^{2}\left(2t\sqrt{J_{L}J_{R}\left[\mathcal{J}_{0}^{2}(\frac{E_{1}}{\omega})+g_{1}(t)\mathcal{J}_{0}(\frac{E_{1}}{\omega})+g_{2}(t)    \right]} \right)\left(\frac{J_{R}}{J_{L}}\right)^m,
\end{eqnarray}
where $g_{1}(t)=2\overline{u}(t)/t$ and $g_{2}(t)=(\overline{u}(t)^{2}+\overline{v}(t)^{2})/t^2$ are vanished at long time, and then we have
\begin{eqnarray}
\rho_{m}(t\gg \frac{2\pi}{\omega})&\approx&\mathcal{J}_{m}^{2}\left(2t\sqrt{J_{L}J_{R}}\mathcal{J}_{0}(\frac{E_{1}}{\omega})\right)\left(\frac{J_{R}}{J_{L}}\right)^m. \label{solac}
\end{eqnarray}
In Fig. \ref{figS3}, we take the  ratio $E_{1}/\omega$  as two non-zeros of $\mathcal{J}_{0}(x)$ (a,c) where show the skin effect and  as two zeros of $\mathcal{J}_{0}(x)$ in (b,d) where show the dynamical localization.

\begin{figure*}[t]
\centering \includegraphics[width=18cm]{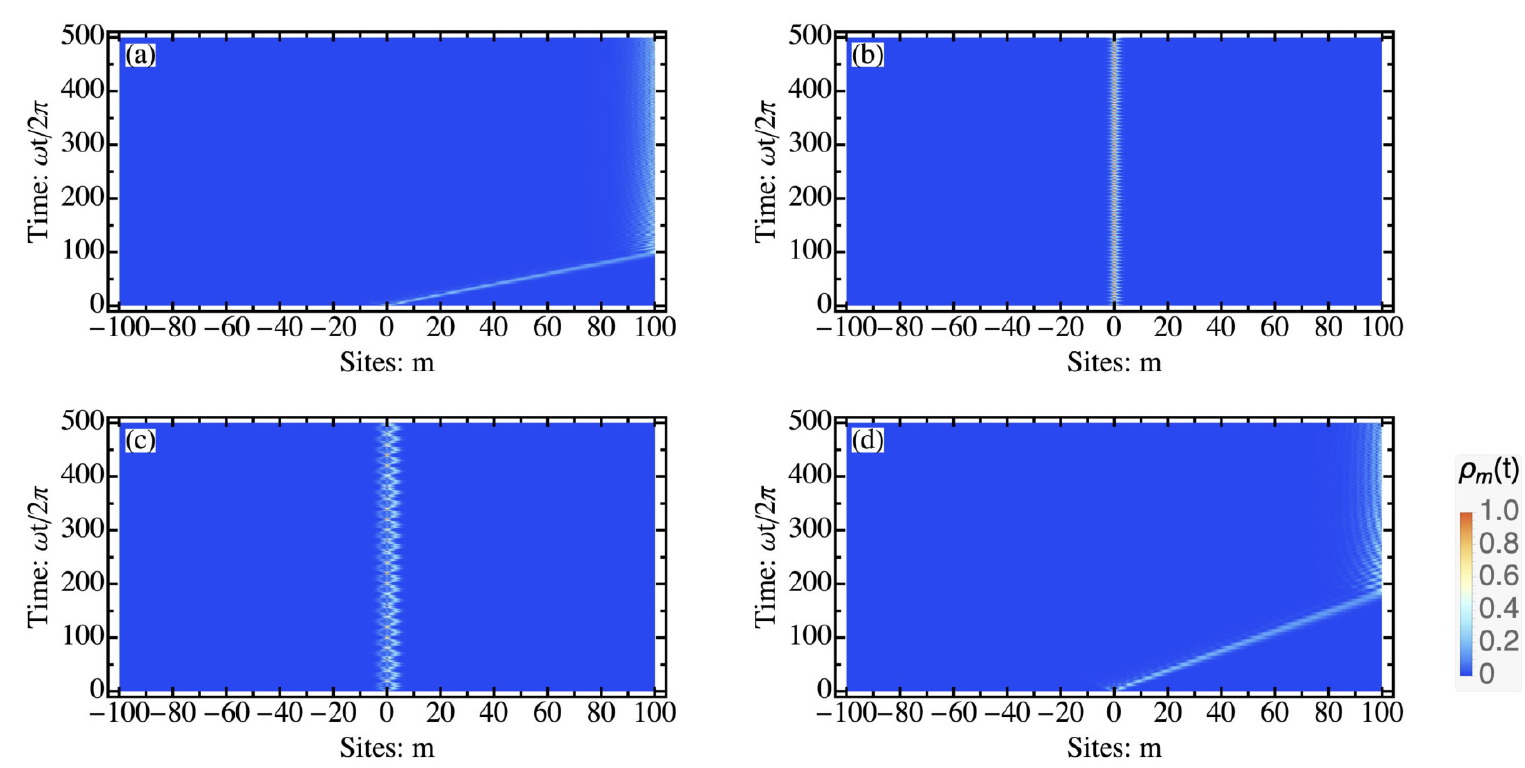}
\caption{ The quantum dynamics for the dc+ac fields with the ratios$~(a):E_{0}/\omega=1.0,~E_{1}/\omega=2.0,~(b):E_{0}/\omega=1.0,~E_{1}/\omega=3.832~(\text{1st zero of~} \mathcal{J}_{1}(x)),~(c):E_{0}/\omega=2.324,~E_{1}/\omega=2.0,~(d):E_{0}/\omega=3.0,~E_{1}/\omega=3.0$. Other parameters are $J_{L}=0.8, ~J_{R}=1.0,~ \omega=1.0$.}
\label{figS4}
\end{figure*}

\subsection{DC+AC electric field}
The results above show that even weakly dc field will push the system to dynamical localization and while the ac field supports the skin effect except the ratio $E_{1}/\omega$ touches the zeros of Bessel function $\mathcal{J}_{0}(x)$. The competition between ac and dc fields in regard of localization will significantly modify the behaviors above.

For this more general case, we firstly rewrite Eq. (\ref{genuv}) as ($\theta=\omega t$)
\begin{eqnarray}
u(\theta)&=&\frac{1}{\omega}\int_{0}^{\omega t}d\theta'\cos\left(\frac{E_{0}}{\omega}\theta'+\frac{E_{1}}{\omega}\sin\theta'\right),\nonumber\\
&=&\frac{1}{\omega}\int_{0}^{\omega t}d\theta'\left[\cos\left(\frac{E_{0}}{\omega}\theta'\right)\cos\left(\frac{E_{1}}{\omega}\sin\theta'\right)-
\sin\left(\frac{E_{0}}{\omega}\theta'\right)\sin\left(\frac{E_{1}}{\omega}\sin\theta'\right)\right],\\
v(\theta)&=&\frac{1}{\omega}\int_{0}^{\omega t}d\theta'\sin\left(\frac{E_{0}}{\omega}\theta'+\frac{E_{1}}{\omega}\sin\theta'\right),\nonumber\\
&=&\frac{1}{\omega}\int_{0}^{\omega t}d\theta'\left[\sin\left(\frac{E_{0}}{\omega}\theta'\right)\cos\left(\frac{E_{1}}{\omega}\sin\theta'\right)+
\cos\left(\frac{E_{0}}{\omega}\theta'\right)\sin\left(\frac{E_{1}}{\omega}\sin\theta'\right)\right],
\end{eqnarray}
Applying the following relations  \cite{BselS},
\begin{eqnarray}
\cos (z \sin \theta)=\mathcal{J}_{0}(z)+2\sum_{k=1}^{+\infty}\mathcal{J}_{2k}(z)\cos(2k\theta),~~~~\sin (z \sin \theta)=2\sum_{k=0}^{+\infty}\mathcal{J}_{2k+1}(z)\sin[(2k+1)\theta],
\end{eqnarray}
and doing the integrating, we arrive at
\begin{eqnarray}
u(t)&=&\frac{\sin (E_{0}t)}{E_{0}}\mathcal{J}_{0}\left(\frac{E_{1}}{\omega}\right)+\sum_{k=1}^{+\infty}(-1)^{k}\mathcal{J}_{k}\left(\frac{E_{1}}{\omega}\right)\left[ \frac{\sin(E_{0}-k\omega)t}{E_{0}-k\omega}+\frac{\sin(E_{0}+k\omega) t}{E_{0}+k\omega} \right],\label{uf1}\\
v(t)&=&\frac{1-\cos (E_{0}t)}{E_{0}}\mathcal{J}_{0}\left(\frac{E_{1}}{\omega}\right)+\frac{2}{\omega}\sum_{k=1}^{+\infty}\mathcal{J}_{2k}\left(\frac{E_{1}}{\omega}\right)\frac{\frac{E_{0}}{\omega}-\frac{E_{0}}{\omega}\cos(E_{0} t)\cos(2k\omega t)-2k\sin(E_{0} t)\sin(2k\omega t)}{(\frac{E_{0}}{\omega})^{2}-4k^2}\nonumber\\
&&+\frac{2}{\omega}\sum_{k=0}^{+\infty}\mathcal{J}_{2k+1}\left(\frac{E_{1}}{\omega}\right)\frac{-(2k+1)+(2k+1)\cos(E_{0}t)\cos\left[(2k+1)\omega t\right]+\frac{E_{0}}{\omega}\sin(E_{0}t)\sin\left((2k+1)\omega t\right]}{(\frac{E_{0}}{\omega})^{2}-(2k+1)^2}.\label{vf1}
\end{eqnarray}
These expressions look very complicated and are difficult in doing the calculations continuously. Even so, we still can do some analysis depends on whether the ratio $E_{0}/\omega$ is an integer or not. If $E_{0}/\omega$ is not an integer, all the terms in Eqs. (\ref{uf1}-\ref{vf1}) are bounded oscillatory functions of time such that the system will localize around the initial states. It means that the dc part of electric field dominate over the ac part and then indicate breaking of the skin effect. In the case of provided $E_{0}/\omega$ is an integer, function $v(t)$ in Eq. (\ref{vf1}) will not deduce any non-bounded functions and will have similar contributions as in the integer case. However, the functions $u(t)$ in Eq. (\ref{uf1}) takes (set $E_{0}=\nu\omega,~\nu\in Z$)
\begin{eqnarray}
u(t)&=&(-1)^{\nu}\mathcal{J}_{\nu}\left(\frac{E_{1}}{\omega}\right)t+\frac{\sin (E_{0}t)}{E_{0}}\mathcal{J}_{0}\left(\frac{E_{1}}{\omega}\right)+\sum_{k=1,k\neq \nu}^{+\infty}(-1)^{k}\mathcal{J}_{k}\left(\frac{E_{1}}{\omega}\right)\left[ \frac{\sin(E_{0}-k\omega)t}{E_{0}-k\omega}+\frac{\sin(E_{0}+k\omega) t}{E_{0}+k\omega} \right],
\end{eqnarray}
and then the first term is a non-bounded and linearly increasing function of time, such that at long time, we have
\begin{eqnarray}
u(t\gg \frac{2\pi}{\omega})\approx(-1)^{\nu}\mathcal{J}_{\nu}\left(\frac{E_{1}}{\omega}\right)t,
\end{eqnarray}
and the probability
\begin{eqnarray}
\rho_{m}(t\gg \frac{2\pi}{\omega})&\approx&\mathcal{J}_{m}^{2}\left(2t\sqrt{J_{L}J_{R}}\mathcal{J}_{\frac{E_{0}}{\omega}}(\frac{E_{1}}{\omega})\right)\left(\frac{J_{R}}{J_{L}}\right)^m,\label{soldcac}
\end{eqnarray}
which is merely replacing the order "0" in Eq. (\ref{solac})  by $E_{0}/\omega$.

As a conclusion, in this "dc+ac" case, the system will be featured in dynamical localization for all non-integer ratio $E_{0}/\omega$ (Fig. \ref{figS4} (c)) and will be dominated by skin effect for all $E_{1}/\omega$ except zeros of $\mathcal{J}_{E_{0}/\omega}(x)$ when the ratio $E_{0}/\omega$ takes integer (Fig. \ref{figS4} (a, b, d)). We show the parameters zone for the skin effect by the solid lines in Fig.3(a) in main text and all the rest zone are for dynamical localization.

\begin{figure*}
\centering
 \includegraphics[width=0.8\textwidth]{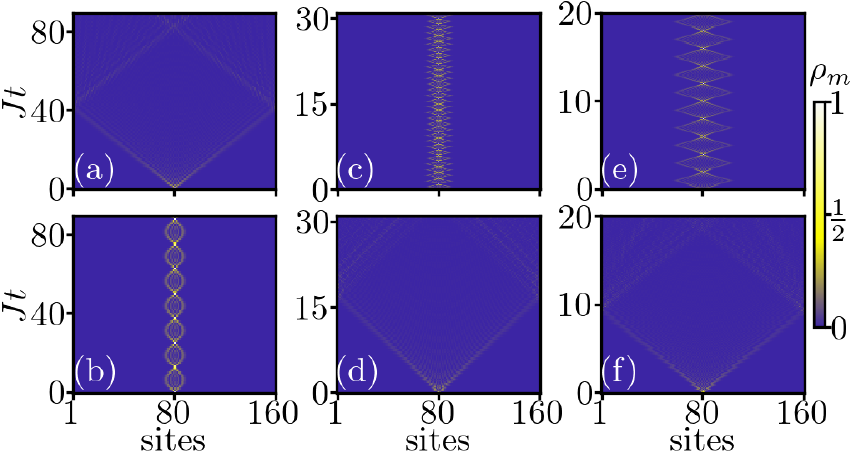}
\caption{\label{bis}
Dynamical evolution of a electron started from the lattice center for the system described by Eq.~(\ref{hami}) with
(a) $E_0=0.005$, $E_1=0$, (b) $E_0=0.5$, $E_1=0$, (c) $E_0=0$, $E_1/\omega=2.405$, which is the first zero point of $\mathcal{J}_0(E_1/\omega)$, (d)
   $E_0=0$, $E_1/\omega=6.1$, (e) $E_0/\omega=0.5$, $E_1/\omega=1.3$, (f) $E_0/\omega=1$, $E_1/\omega=5.7$. Except $J_R$ and $J_L$, (a),(b) and Fig.1 (e),(f) of the main text have same parameters, (c),(d) and Fig.2 (c),(d) of the main text have same parameters, (e),(f) and Fig.3 (b),(d) of the main text have same parameters. Here $J_R=J_L=1$.}
\end{figure*}

\subsection{Hermitian case}
To understand the physical picture mentioned in the main text more intuitively, we here provide the time evolution of the corresponding Hermitian case, i.e., $J_R=J_L$, as shown in Fig. \ref{bis}. Figs. \ref{bis} (a),(b) show the dc case, whose parameters are same with Figs.1 (e),(f) of the main text. The previous discussions about the dc case can be directly applied to the $J_R=J_L$ case, as shown in Fig. \ref{bis} (b). Although both Fig. \ref{bis}(b) and Fig.1(f) of the main text show the Stark localization, there also exist the differences. From Fig. \ref{bis} (b), the particle oscillate around the initial position, but when $J_R>J_L$, the oscillation of the particle is at the right-hand side of the initial position [Fig.1(f)]. Comparing Fig. \ref{bis}(c), Fig.\ref{bis}(e) and Fig.2(c), Fig.3 (b) of the main text, we can also clearly see the phenomenon that the  oscillation center move to the right of the initial position when changing $J_L=J_R$ to $J_L<J_R$.

When $E_0$ is decreased, as discussed below, the oscillation amplitude will increase. For a finite system, when the oscillation amplitude is larger than the system size, the particle can arrive at the boundary, as shown in Fig. \ref{bis} (b). When $J_R>J_L$, the particle
will arrive at the boundary and then stay there ever since [Fig.1(e)], which corresponds to the existence of the NHSE. The competition between the oscillation amplitude and the system size induce the phenomenon of the size-dependent NHSE. Comparing Fig. \ref{bis}(d), Fig.\ref{bis}(f) and Fig.2(d), Fig.3 (d) of the main text, it can be seen that the extended states become the skin modes when changing $J_L=J_R$ to $J_L<J_R$.

%\bibliography{Citations}
%\end{widetext}
\section{III. Finite size effect for the pure dc case}
In this section, we investigate the finite size effect in the pure dc field case. According to the inequality proven by Paris~\cite{Paris} for Bessel function,
the probability $\rho_m(t)$ is bounded from above for big enough $m$,
\begin{equation}
    \rho_m(t)
    \le\mathcal{J}^2_{m-n_0}(m-n_0)f_\chi^{2(m-n_0)}(x_t).
    \label{eq_rho_m_upperB}
\end{equation}
Here $f_\chi(x_t)=x_te^{1-\chi{}x_t}$, $\chi=\sqrt{J_L/J_R}<1$, and
$x_t=\frac{4J_R}{(m-n_0)|E_0|}\left|\sin\frac{E_0t}{2}\right|$. The inequality holds
if $4\sqrt{J_LJ_R}\le(m-n_0)|E_0|$. $f_\chi(x_t)$ is monotonically increasing
and bounded $0\le{}f_\chi(x_t)\le1/\chi$, for $0\le{}x_t\le1/\chi$. One can calculate
the derivation of $f_\chi$ to verify it.
\begin{equation}
    f'_{\chi}(x_t)
    = e^{1-\chi{}x_t}-\chi{}x_te^{1-\chi{}x_t}
    = e^{1-\chi{}x_t}\left(1-\chi{}x_t\right).
    \label{eq_f_chi_1st_d}
\end{equation}
And $f_\chi(x_t=1)=e^{1-\chi}>1$ since $\chi<1$.
There exists
$0\le{}x_\star<1$ such that $f_\chi(x_\star)=1$ and $0\le{}f_\chi(x_t)<1$ if
$0\le{}x_t<x_\star$.
Note that $x_\star$ depends only on $J_L$ and $J_R$, and satisfies $x_\star e^{1-\sqrt{J_L/J_R}x_\star}=1$.
We also know $|\mathcal{J}_n(\bullet)|\le1$~\cite{Olver}.
Thus
\begin{eqnarray}
    \rho_m(t)
    &\le&
    \mathcal{J}^2_{m-n_0}(m-n_0)f_\chi^{2(m-n_0)}\left(\frac{4J_R}{(m-n_0)|E_0|}\left|\sin\frac{E_0t}{2}\right|\right)
    \nonumber\\
    &\le&
    \mathcal{J}^2_{m-n_0}(m-n_0)\left[\underbrace{f_\chi\left(\frac{4J_R}{(m-n_0)|E_0|}\right)}_{<1}\right]^{2(m-n_0)}
    \nonumber\\
    &\le&
    \left[\underbrace{f_\chi\left(\frac{4J_R}{(m-n_0)|E_0|}\right)}_{<1}\right]^{2(m-n_0)}.
\end{eqnarray}
As $m$ increases, $f_\chi\left({4J_R}/{(m-n_0)|E_0|}\right)$ would decrease further.
Thus the decay rate of $\rho_m$ is faster than any exponential decay rate
as $m$ increases.
Hence $\rho_m(t)$ would be insignificant if $4J_R<(m-n_0)|E_0|x_\star$.
It is well known that $\mathcal{J}_{-n}(\bullet)=(-1)^n\mathcal{J}_n(\bullet)$.
For two sites $2n_0-m$ (left side $<n_0$) and $m$ (right side $>n_0$)
with equal distance to $n_0$
\begin{equation}
    \frac{\rho_{2n_0-m}(t)}{\rho_{m}(t)}
    \approx
    \frac{ \mathcal{J}_{n_0-m}^{2}\left(2t\sqrt{J_{L}J_{R}}
    \mathcal{J}_{0}(\frac{E_{1}}{\omega})\right)\left(\frac{J_{R}}{J_{L}}\right)^{n_0-m}}
     { \mathcal{J}_{m-n_0}^{2}\left(2t\sqrt{J_{L}J_{R}}
    \mathcal{J}_{0}(\frac{E_{1}}{\omega})\right)\left(\frac{J_{R}}{J_{L}}\right)^{m-n_0}}
    =\left(\frac{J_{R}}{J_{L}}\right)^{-2(m-n_0)}.
    \label{eq_rho_m_spoints}
\end{equation}
Adding the fact that $J_R>J_L$, $\rho_{2n_0-m}(t)$ would be even less significant,
if $m>n_0$ and $4J_R<(m-n_0)|E_0|x_\star$.
Therefore, in an infinite lattice, an electron
would oscillate around its initial position $n_0$ within a range of
$4J_R/|E_0|x_\star$.
\section{IV. Dynamical behavior of a particle on non-reciprocal tight-binding model with non-Hermitian skin effect}
In this section, we take the case without external field as an example to discuss the dynamical behavior of a particle. When there is no electric field, Eq.~\eqref{eq_rho_m_upperB} and Eq.~\eqref{eq_rho_m_spoints}
would still be valid except $x_t=2tJ_R/(m-n_0)$. Eq.~\eqref{eq_rho_m_spoints}
tells us that the electron would always favor the right-hand side of $n_0$.
 Further, from Eq.~\eqref{eq_f_chi_1st_d} we know that
$f_\chi(x_t)$ increases from 0 to $f^{\mathrm{max}}_\chi=\sqrt{J_R/J_L}$ as
$x_t$ increases from 0 to $1/\chi$. Hence $f_\chi(x_t)<1$ and thus
the upper bound of $\rho_m$ (see Eq.~\eqref{eq_rho_m_upperB}) would be very small, when $x_t<x_\star$.
Here $f_\chi(x_\star)=1$ and $x_\star<1$ as mentioned in the previous section.
Hence $\rho_m$ would be significant only if $x_t{\ge}x_\star$, namely
\begin{equation}
    t\ge\frac{(m-n_0)x_\star}{2J_R}.
\end{equation}
Therefore, the electron reaches site $m$ around $t_a=\frac{(m-n_0)x_\star}{2J_R}$
and continues hopping rightward. Now we consider the stable probability in a long time after the particle passing.
As discussed above, in the absence of external fields, $\rho_{m}(t)=\mathcal{J}_{m-n_0}^{2}\left(2t\sqrt{J_{L}J_{R}}\right)(J_{R}/J_{L})^{m-n_0}$, where $2t\sqrt{J_{L}J_{R}}$ is linearly increasing to infinity with increasing time, and thus $\mathcal{J}_{m-n_0}(2t\sqrt{J_{L}J_{R}})$ tends to 0.
We consider $\rho_{m}/\rho_{m-1}$ when $t\rightarrow\infty$. For convenience, we consider the long time averaged behavior after the particle has passed the position $m$ for a long time.
We notice that $\frac{\int^{\infty}_{t\gg t_a}dt\mathcal{J}_{m-n_0}^{2}\left(2t\sqrt{J_{L}J_{R}}\right)}{\int^{\infty}_{t\gg t_a}dt\mathcal{J}_{m-1-n_0}^{2}\left(2t\sqrt{J_{L}J_{R}}\right)}=1$ and so $\rho_{m}/\rho_{m-1}=J_{R}/J_{L}$. Thus, for a system with non-Hermitian skin effect, the probability distributions of the region that the particle has passed through are not ergodic, which is obviously different from the system without skin effect.
%%%%%%%%%%%%%%%%%%%%%%%%%%%%%%%%%%%%%%%%%%%%%%%%%%%%

%%%%%%%%%%%%%%%%%%%%%%%%%%%%%%%%%%%%%%%%%%%%%%%%%%%%%
\end{document}